% version 3
\input harvmac
\input epsf

\overfullrule=0pt
\abovedisplayskip=12pt plus 3pt minus 1pt
\belowdisplayskip=12pt plus 3pt minus 1pt
%macros
%
\def\tilde{\widetilde}
\def\bar{\overline}

\def\tL{{\tilde L}}

\def\bigone{\hbox{1\kern -.23em {\rm l}}}
\def\ZZ{\hbox{\zfont Z\kern-.4emZ}}
\def\half{{\litfont {1 \over 2}}}

\font\litfont=cmr6

\def\ddb#1{{${\rm D}#1-{\bar{\rm D}#1}$}}
\def\Dbar{${\bar {\rm D}}$}

\def\cF{{\cal F}}
\def\cV{{\cal V}}
\def\Ei{\,{\rm Ei}\,}

\nref\bansuss{T. Banks and L. Susskind, {\it ``Brane--Antibrane
Forces''}, hep-th/9511194.}
\nref\naraintach{E. Gava, K.S. Narain and M.H. Sarmadi, {\it ``On the 
Bound States of p-Branes and (p+2)-Branes''}, hep-th/9704006;
Nucl. Phys. {\bf B504} (1997) 214.}
\nref\sentach{A. Sen, {\it ``Tachyon Condensation on the Brane
Anti-Brane System''}, hep-th/9805170; JHEP {\bf 08} (1998) 012.}
\nref\sred{M. Srednicki, {\it ``IIB or Not IIB''}, hep-th/9807138;
JHEP {\bf 08} (1998) 005.}
\nref\senso{A. Sen, {\it ``SO(32) spinors of type I and other
solitons on brane - anti-brane pair''}, hep-th/9808141;
JHEP {\bf 09} (1998) 023.}
\nref\witk{E. Witten, {\it ``D-branes and K-theory''},
hep-th/9810188; JHEP {\bf 12} (1998) 019.}
\nref\hork{P. Horava, {\it ``Type IIA D-Branes, K-Theory and Matrix
Theory''}, hep-th/9812135; Adv. Theor. Math. Phys. {\bf 2} (1999)
1373.}
\nref\piljin{P. Yi, {\it ``Membranes from Five-Branes and Fundamental
Strings from Dp-Branes''}, hep-th/9901159; 
Nucl. Phys. {\bf B550} (1999) 214.}
\nref\sendesc{A. Sen, {\it ``Descent Relations Among Bosonic
D-branes''}, hep-th/9902105; Int. J. Mod. Phys. A14 (1999) 4061.}
\nref\awata{H. Awata, S. Hirano and Y. Hyakutake, {\it ``Tachyon Condensation
and Graviton Production in Matrix Theory''}, hep-th/9902158.}
\nref\berget{O. Bergman, E. Gimon and P. Horava, {\it ``Brane Transfer
Operations and T-duality of Non-BPS States''}, hep-th/9902160;
JHEP {\bf 04} (1999) 010.}
\nref\pesando{I. Pesando, {\it ``On the Effective Potential of the
Dp--anti-Dp System in Type II Theories''}, hep-th/9902181;
Mod. Phys. Lett. {\bf A14} (1999) 1545.}
\nref\frau{M. Frau, L. Gallot, A. Lerda and P. Strigazzi, 
{\it ``Stable Non-BPS D-branes in Type I String Theory''},
hep-th/9903123; Nucl. Phys. {\bf B564} (2000) 60.}
\nref\nakwoo{N. Kim, S.-J. Rey and J.-T. Yee, {\it ``Stable 
Non-BPS Membranes on M(atrix) Orientifold''}, hep-th/9903129; JHEP
{\bf 04} (1999) 003.}
\nref\dmfrac{K. Dasgupta and S. Mukhi, {\it ``Brane Constructions, 
Fractional Branes and Anti-de Sitter Domain Walls''}, hep-th/9904131;
JHEP {\bf 07} (1999) 008.}
\nref\dealwis{S.P. de Alwis, {\it ``Tachyon Condensation in Rotated
Brane Configurations''}, hep-th/9905080; Phys. Lett. {\bf B461}
(1999) 329.}
\nref\kenwilk{C. Kennedy and A. Wilkins, {\it ``Ramond-Ramond
Couplings on Brane-Antibrane Systems''}, hep-th/9905195;
Phys. Lett. {\bf B464} (1999) 206.}
\nref\aldauranga{G. Aldazabal and A.M. Uranga, {\it ``Tachyon Free
Nonsupersymmetric Type IIB Orientifolds Via Brane-AntiBrane 
Systems''}, hep-th/9908072; JHEP {\bf 10} (1999) 024.}
\nref\gabstef{M. Gaberdiel and B. Stefanski, {\it ``Dirichlet Branes on
Orbifolds''}, hep-th/9910109.}
\nref\joysen{J. Majumder and A. Sen, {\it `` `Blowing Up' D-Branes 
on Nonsupersymmetric Cycles''}, hep-th/9906109; JHEP {\bf 09} (1999)
004.} 
\nref\angelant{C. Angelantonj, {\it ``Non-supersymmetric Open
String Vacua''}, hep-th/9907054.}
\nref\antoniad{I. Antoniadis, E. Dudas and A. Sagnotti, 
{\it ``Brane Supersymmetry Breaking''}, hep-th/9908023;
Phys. Lett. {\bf B464} (1999) 38.}
\nref\gabsen{M. Gaberdiel and A. Sen, {\it ``Nonsupersymmetric
D-Brane Configurations with Bose-Fermi Degenerate Open String
Spectrum''}, hep-th/9908060; JHEP {\bf 11} (1999) 008.}
\nref\youm{D. Youm, {\it ``Delocalized Supergravity Solutions
for Brane/Anti-brane Systems and Their Bound States''},
hep-th/9908182.}
\nref\senworld{A. Sen, {\it ``Supersymmetric World-Volume Action For
Non-BPS D-Branes''}, hep-th/9909062; JHEP {\bf 10} (1999) 008.} 
\nref\tatar{M. Mihailescu, K. Oh and R. Tatar, {\it ``Non-BPS Branes 
on a Calabi-Yau Threefold and Bose-Fermi Degeneracy''}, hep-th/9910249.}
\nref\houart{L. Houart and Y. Lozano, {\it ``Type II
Branes from Brane-antibrane in M-theory''}, hep-th/9910266.}
\nref\angelanttwo{C. Angelantonj, I. Antoniadis, G. D'Appollonio,
E. Dudas and A. Sagnotti, {\it ``Type I Vacua with
Brane Supersymmetry Breaking''}, hep-th/9911081.}
\nref\russcr{R. Russo and C.A. Scrucca, {\it ``On the
Effective Action of Stable Non-BPS Branes''}, hep-th/9912090.}
\nref\sentachpot{A. Sen, {\it ``Universality of the Tachyon
Potential''}, hep-th/9911116; JHEP {\bf 12} (1999) 027.}
\nref\augusto{A. Sagnotti, {\it ``Open String Models with
Broken Supersymmetry''}, hep-th/0001077.}
\nref\berhoryi{O. Bergman, K. Hori and P. Yi, {\it 
``Confinement on the Brane''}, hep-th/0002223.}
\lref\senzwie{A. Sen and B. Zwiebach, {\it ``Tachyon
Condensation in String Field Theory''}, hep-th/9912249;
JHEP {\bf 03} (2000) 002.}
\lref\berko{N. Berkovits, {\it ``The Tachyon Potential in
Open Neveu-Schwarz String Field Theory''}, hep-th/0001084.}
\lref\watione{W. Taylor, {\it ``D-brane Effective Field
Theory from String Field Theory''}, hep-th/0001201.}
\lref\harkraus{J.A. Harvey and P. Kraus, {\it ``D-branes as 
Unstable Lumps in Bosonic Open String Field Theory''}, 
hep-th/0002117.}
\lref\berksenzwie{N. Berkovits, A. Sen and B. Zwiebach, {\it ``Tachyon 
Condensation in Superstring Field Theory''}, hep-th/0002211.}
\lref\moetaylor{N. Moeller and W. Taylor, {\it ``Level
Truncation and the Tachyon in Open Bosonic String Field Theory''},
hep-th/0002237.}
\lref\harkuma{J.A. Harvey, D. Kutasov and E. Martinec, {\it ``On 
the Relevance of Tachyons''}, hep-th/0003101.}
\nref\senreview{A. Sen, {\it ``Non-BPS States and Branes in String
Theory''}, hep-th/9904207.}
\nref\lerdarev{A. Lerda and R. Russo, {\it ``Stable Non-BPS
States in String Theory: A Pedagogical Review''}, hep-th/9905006.}
\nref\bergabrev{O. Bergman and M. Gaberdiel, {\it ``Non-BPS Dirichlet
Branes''}, hep-th/9908126.}
\nref\schwarzrev{J. H. Schwarz, {\it ``TASI Lectures on Non-BPS 
D-brane systems''}, hep-th/9908144.}
\lref\mst{S. Mukhi, N.V. Suryanarayana and D. Tong, {\it
``Brane-Antibrane Constructions''}, hep-th/0001066.}
\lref\dmconif{K. Dasgupta and S. Mukhi, {\it ``Brane Constructions,
Conifolds and M-Theory''}, hep-th/9811139; Nucl. Phys. {\bf B551} 
(1999) 204.}
\lref\urangaconif{A. Uranga, {\it ``Brane Configurations for Branes at
Conifolds''}, hep-th/9811004; JHEP {\bf 01} (1999) 022.}
\lref\gns{S. Gubser, N. Nekrasov and S. Shatashvili, {\it 
``Generalized Conifolds and 4-dimensional N=1 Superconformal Field
Theory''}, hep-th/9811230; JHEP {\bf 05} (1999) 003.}
\lref\witfourd{E. Witten, {\it ``Solutions of Four Dimensional Field
Theories Via M-theory''}, hep-th/9703166; Nucl. Phys. {\bf B500}
(1997) 3.}
\lref\hanwit{A. Hanany and E. Witten, {\it ``Type IIB Superstrings, 
BPS Monopoles, and Three-dimensional Gauge Dynamics''},
hep-th/9611230; Nucl. Phys. {\bf B492} (1997) 152.}
\lref\dougmoore{M. Douglas and G. Moore, {\it ``D Branes, Quivers and ALE
Instantons''}, hep-th/9603167.}
\lref\nsdual{H. Ooguri and C. Vafa, {\it ``Two-Dimensional Black Hole
and Singularities of CY Manifolds''}, hep-th/9511164; Nucl. Phys. 
{\bf B463} (1996) 55\semi
B. Andreas, G. Curio and D. L\"ust, {\it ``The Neveu-Schwarz 
Five-Brane and its Dual Geometries''}, hep-th/9807008; 
JHEP {\bf 10} (1998) 022\semi
A. Karch, D. L\"ust and D. Smith, {\it ``Equivalence of 
Geometric Engineering and Hanany-Witten via Fractional Branes''},
hep-th/9803232; Nucl. Phys. {\bf B533} (1998) 348.}
\lref\kehagias{A. Kehagias, {\it ``New Type IIB Vacua and Their
F-theory Interpretation''}, Phys. Lett. {\bf B435} (1998) 337;
hep-th/9805131.}
\lref\klebnek{I. Klebanov and N. Nekrasov, {\it ``Gravity Duals of
Fractional Branes and Logarithmic RG Flow''}, hep-th/9911096.}
\lref\jpp{C.V. Johnson, A.W. Peet and J. Polchinski,
{\it ``Gauge Theory and the Excision of Repulson Singularities''},
hep-th/9911161.}
\lref\klebtseyt{I. Klebanov and A.A. Tseytlin, {\it ``Gravity Duals of 
Supersymmetric $SU(N)\times SU(N+M)$ Gauge Theories''},
hep-th/0002159.}
\lref\ohtatar{K. Oh and R. Tatar, {\it ``Renormalization Group 
Flows on D3 branes at an Orbifolded Conifold''}, hep-th/0003183.}

{\nopagenumbers
\Title{\vbox{
\hbox{hep-th/0003219}
\hbox{TIFR/TH/00-12}}}
{\vtop{\centerline{A Stable Non-BPS Configuration From}
\medskip
\centerline{Intersecting Branes and Antibranes}}}
\centerline{{Sunil Mukhi}\footnote{}{E-mail: mukhi@tifr.res.in,
nemani@tifr.res.in} and {Nemani V. Suryanarayana}}
\vskip 8pt
\centerline{\it Tata Institute of Fundamental Research,}
\centerline{\it Homi Bhabha Rd, Mumbai 400 005, India}

\vskip 3truecm
\centerline{\bf ABSTRACT}
We describe a tachyon-free stable non-BPS brane configuration in type
IIA string theory. The configuration is an elliptic model involving
rotated NS5 branes, D4 branes and anti-D4 branes, and is dual to a
fractional brane-antibrane pair placed at a conifold singularity. This
configuration exhibits an interesting behaviour as we vary the radius
of the compact direction. Below a critical radius the D4 and anti-D4
branes are aligned, but as the radius increases above the critical
value the potential between them develops a minimum away from zero.
This signals a phase transition to a configuration with finitely
separated branes.
\vfill
\Date{March 2000}
\eject}
\ftno=0

\noindent{\bf Introduction and Review}

Much has been learned in recent times about the physics of
brane-antibrane pairs and non-BPS branes in superstring
theory\refs{\bansuss-\berhoryi}. Parallel, infinitely extended pairs
attract each other, and can annihilate into the vacuum by a process of
tachyon condensation into a constant minimum. An analogous decay
process takes place for single or multiple non-BPS
branes. Condensation of the tachyon as a kink, vortex or more general
soliton is associated to brane-antibrane annihilation into branes of
lower dimension.

The above considerations have been extended to backgrounds with lower
supersymmetry (orientifolds, orbifolds and smooth Calabi-Yau
manifolds), where one finds new phenomena including the existence of
stable, non-BPS branes. As parameters of the background are varied,
there can also be phase transitions between qualitatively different
configurations. The reader may consult the reviews in
Refs.\refs{\senreview-\schwarzrev}. 

A different direction, explored in Ref.\refs\mst, is to consider
non-BPS configurations of intersecting branes and antibranes in the
fully supersymmetric type II spacetime background. Here one encounters 
novel phenomena including both attractive and repulsive interactions
among branes and antibranes. Such configurations could be useful to
study non-supersymmetric field theories, and also to understand better 
the basic underlying structure of superstring theory.

In the present note we examine a variant of a configuration of
``adjacent brane-antibrane pairs'' that was discussed in
Ref.\refs\mst. Let us describe the original configuration. In type IIA
theory we start with a pair of parallel NS5-branes extended along
$x^1,x^2,x^3,x^4,x^5$ and separated along $x^6$. The $x^6$ direction
is compact, with circumference $2L$. Now stretch a $D4$-brane along
$x^6$ from the first NS5-brane to the second, and a \Dbar4-brane along
$x^6$ from the second NS5-brane to the first (Fig.1).
\bigskip

\centerline{\epsfbox{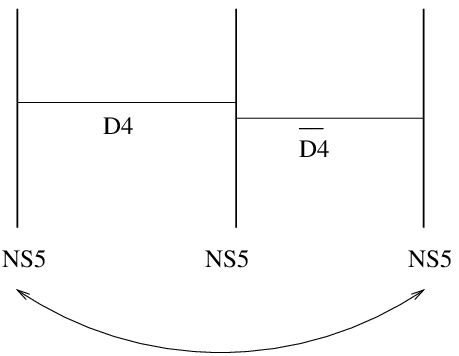}}\nobreak
\centerline{Fig.1: Adjacent D4 and \Dbar4 between parallel branes.}

\bigskip
In Ref.\refs\mst\ it was argued that the D4-brane and
\Dbar4-brane exert a net repulsive force on each other, with the
result that the configuration is unstable. From the point of view of
the field theory on the NS5 world volume, this repulsion is
essentially due to the fact that the D4 and \Dbar4 end on the NS5
brane from opposite sides, and their ends are charged 3-branes in the
NS5 world volume. If we dimensionally reduce everything over these
three directions, then the ends become vortices living on the reduced
NS5 world volume. These vortices carry the same charge under the gauge
field, hence they repel, and since the configuration is
non-supersymmetric there is no reason to expect that this repulsion is
cancelled by exchange of other massless fields. Because the NS5-branes
are parallel, the repelling D4- and \Dbar4-branes can run away from
each other to infinity.

This instability can also be understood in the T-dual picture, where
the D4- and \Dbar4-branes are actually two types of fractional branes
(denoted $1_f$ and $\bar{1}'_f$ respectively in \refs\mst) at a $Z_2$
ALE singularity. These two fractional branes repel, as they each carry
a full unit of twisted RR charge. There is also an attraction due to
the fractional untwisted RR charge, but an explicit computation of the
amplitude using orbifold techniques\refs\mst\ reveals that the
repulsive force dominates.

The variant of this configuration that we will describe in the next
section involves rotating the NS5-branes. This converts the T-dual ALE
space into a conifold\refs{\urangaconif,\dmconif}, hence this brane
construction is now T-dual to fractional branes at a conifold, for
which we cannot use orbifold techniques to compute the force. We will
analyse the model using some observations in
Refs.\refs{\urangaconif,\dmconif,\gns,\dmfrac,\mst}\ and argue that
this time a stable non-BPS configuration is obtained.
\bigskip

\noindent{\bf A Stable Configuration}

Consider the following brane configuration in Type IIA: an NS5-brane
filling $x^1, x^2, x^3, x^4, x^5$ and located at $(x^8,x^9)=(0,0)$,
and another NS5-brane (denoted by NS5') filling $x^1, x^2, x^3, x^8,
x^9$ and located at $(x^4,x^5)=(0,0)$. The two branes are placed at the
same point in the $x^7$ direction and are separated along the compact
$x^6$ direction of circumference $R_6$. 

Suspend a D4-brane between NS5-NS5' in one of the two intervals and a
\Dbar4 in the other, so that the 4-branes extend along $x^1, x^2,
x^3$ and $x^6$ (Fig.2). This configuration breaks all the supersymmetries of
Type IIA, though each of the D4 and \Dbar4, together with the
NS5-branes, separately preserves some supersymmetry.
\bigskip

\centerline{\epsfbox{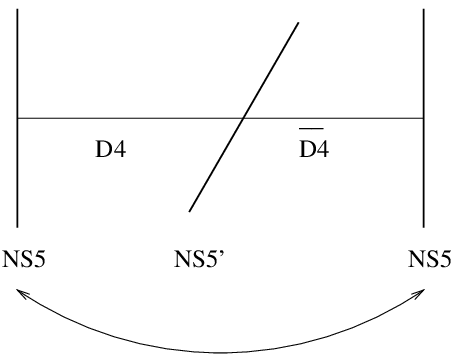}}\nobreak
\centerline{Fig.2: Adjacent D4 and \Dbar4 between rotated branes.}

\bigskip
Unlike the case discussed in the previous section, here the
configuration of NS5-branes is no longer dual to an ALE space but
rather to a conifold\refs{\urangaconif,\dmconif}. Nevertheless, one
can still argue that the stretched D4- and
\Dbar4-branes repel. In the case of parallel
NS5-branes, the repulsion was identified as coming from like charges
carried by the ends of the 4-branes. In this picture the repulsive
effect is localised on each NS5-brane separately, hence introducing a
relative rotation should not matter. Moreover, from the string theory
point of view, the repulsion between the D4 and
\Dbar4 is obtained by calculating a closed-string tree amplitude
(cylinder amplitude). Translated into the open-string channel, this is
a one-loop open-string amplitude. This amplitude depends only on the
spectrum obtained by quantizing the open strings connecting D4 and
\Dbar4 across any one NS5 brane. This again suggests that the force is
localized near one NS5 brane at a time\foot{This feature of open
strings across NS5-branes was exploited in
Refs.\refs{\urangaconif,\dmconif}\ to obtain the spectrum of the gauge
theory living on related (BPS) brane configurations.}. Using these
arguments, we conclude that the D4- and \Dbar4-branes in Fig.2 repel
each other, and that the repulsion is the same as that between
adjacent D4- and \Dbar4-branes when the NS5-branes are not rotated
with respect to each other.

With rotated NS5-branes, the important difference is that the the
D4-branes no longer have moduli to move away from each other. As they
move with their ends on the NS5-branes, the D4-branes get stretched.
In the process their effective 3-brane tension increases, providing a
restoring force for the repelling ends of the adjacent \ddb4. Thus one
can expect a configuration in which the repulsive force and the
restoring force due to the increased tension of the adjacent \ddb4\
pair exactly cancel, giving rise to a configuration that is stable
at least under small perturbations.

In fact, as we now show explicitly, such a stable configuration exists
for some range of values of the circumference $R_6$. For simplicity,
let us assume that the NS5 and NS5'-branes are located at
diametrically opposite points on the compact $x^6$ direction, with the
separation between them being $L=\half R_6$. With this, and the fact
that the branes are rotated at 90 degrees to each other, there is a
high degree of symmetry in the problem.  If we let $r$ be the
displacement of the end of the D4 brane from the origin in the $x^4$
(or $x^5$) direction, then the
\Dbar4-brane will also be displaced by an equal amount $r$, and the
displacement of the other ends of the 4-branes along $x^8$ (or $x^9$)
will also be $r$ (Fig.3). 
\bigskip
\centerline{\epsfbox{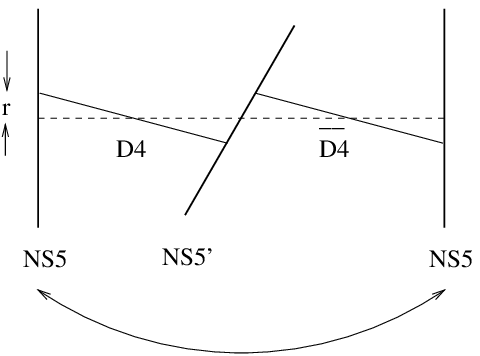}}\nobreak
\centerline{Fig.3: Equilibrium configuration after displacement of 
4-branes.}

\bigskip

With the above data, the net tension of the stretched D4(\Dbar4) is
$\cV\, T_4 \sqrt{L^{2}+2r^{2}}$ where $\cV$ is the (infinite) 3-volume
of the $(x^1,x^2,x^3)$ directions and $T_4= {1\over g_s (2\pi)^4}$ is
the tension of a BPS D4-brane. The contribution from the repulsion
between the ends of D4- and \Dbar4-branes on an NS5-brane to the
energy of the system is given by\refs\mst:
\eqn\repulsion{
{\cV \over 16(2\pi)^4}\int_{0}^{\infty}{dt \over t^3}
e^{-{2X^{2}t\over \pi}}\cF(q)} 
Here $X=2r$ is the separation of D4 and \Dbar4 along the NS5 brane,
and $q=\exp(-\pi t)$. The function $\cF(q)$ is given by 
\eqn\fqdef{
\cF(q)={f_4(q)^8\over f_1(q)^8}
\left[1 - 4{f_1(q)^4 f_3(q)^4 \over
f_2(q)^4 f_4(q)^4}\right] }
where the $f_i(q)$ are defined as:
\eqn\fourellip{\eqalign{
&f_1(q) = q^{1\over 12}\prod_{n=1}^{\infty}(1 - q^{2n})\qquad\qquad
f_2(q) = \sqrt {2} q^{1\over12}\prod_{n=1}^{\infty}(1 + q^{2n})\cr
&f_3(q)= q^{-{1\over24}}\prod_{n=1}^{\infty}(1 + q^{2n-1})\qquad ~
f_4(q) = q^{-{1\over24}}\prod_{n=1}^{\infty}(1 - q^{2n-1})\cr }}
Dropping the common factor $\cV$, the total potential energy of the
system of branes in Fig.3 is
\eqn\energy{
\eqalign{
V(r) &=  {1\over g_s (2\pi)^4}\sqrt{L^{2}+2r^{2}} -
{1 \over 16(2\pi)^4}\int_{0}^{\infty}{dt \over t^3}
e^{-{8r^{2}t\over \pi}} \cF(q)\cr
&= V^{(1)}(r) + V^{(2)}(r)\cr}}
This expression can be minimized to get the condition for the
equilibrium value for $r$.

The second term in Eqn.\energy\ is somewhat complicated. So we first
analyze it in two different limits and extract some physical
information. In the $r\rightarrow \infty$ limit, the most significant
contribution comes from the $t \rightarrow 0$ behaviour of $\cF(q)$.
\eqn\largex{
\cF(e^{-\pi t}) \rightarrow -16t^2 \quad \hbox{as} \quad t \rightarrow 0}
Therefore we have 
\eqn\largxint{
V^{(2)}(r) = -{1 \over 16(2\pi)^4}\int_{0}^{\infty}{dt \over t^3}
e^{-{8r^{2}t\over \pi}}(-16t^2),\qquad r\gg 1}
(we are measuring distances in units of $\sqrt{\alpha'}$). 
This integral diverges logarithmically because of the behaviour of the
integrand in the $t \rightarrow 0$ limit. To extract the behaviour of
this quantity as a function of $r$, let us regulate it by putting a
cut-off $\epsilon$ for the lower limit of the integration variable $t$,
and then take $\epsilon\rightarrow 0$. Thus Eqn.\largxint\ becomes
\eqn\lrgxvlu{
V^{(2)}(r) =
{1\over{(2\pi)^4}}\left[-{\gamma}
-\log({{8r^{2}{\epsilon}}\over{\pi}}) - \sum_{n =
1}^{\infty}{(-1)^{n}
({8r^{2}{\epsilon}\over{\pi}})^{n}\over{n.n!}}\right],\qquad r\gg 1}
where $\gamma$ is the Euler constant. In the limit $\epsilon
\rightarrow 0$ the third term vanishes and we are left with a
potential of the form:
\eqn\lgxpot{
V^{(2)}(r) = A - B\, \log(r), \qquad r\gg 1}
where $A$ is an infinite constant and $B =
{2\over{(2\pi)^4}}$. From this the contribution of the second
term in Eq.\energy\ to the force between the two
$D4$-branes, given by $-{dV^{(2)}\over dr}$, is
\eqn\fifty{
F^{(2)}(r) = {B \over r}, \qquad r\gg 1}
Thus in the large-separation limit this contribution to the force
between two 4-brane segments is repulsive, as expected.

Now let us look at the behaviour of this contribution for small values of
$r$.  In this limit we can expand the the exponential in Eqn.\energy\
in powers of $r$ to get
\eqn\smlxpot{
V^{(2)}(r) = C -  D r^2, \qquad r\ll 1}
where
\eqn\cdvalues{
\eqalign{
C &= {{1 \over
{8(2\pi)^4}}\int_0^{\infty}{dt \over {t^3}}\cF(q)} \cr D & = -
{{1 \over {(2 \pi)^5}}}\int_0^{\infty}{dt \over
{t^2}}\cF(q)}}
Notice that $C$ is a divergent integral whereas $D$ is convergent. $D$
is also positive because $\cF(q)$ is negative all through the range
of integration. From Eqn.\smlxpot\ the small-$r$ behaviour of the
force turns out to be
\eqn\fzero{
F^{(2)}(r) = 2 D r, \qquad r\ll 1}
which is also repulsive, as expected. From Eqn.\energy, the restoring
force is:
\eqn\rstfor{
F^{(1)}(r) = -{d V^{(1)}\over dr} = - 
{1\over g_s (2\pi)^4}{2r \over \sqrt{2L^{2} + {(2r)}^{2}}}}
which is attractive as explained above. The strength of attraction
depends on the value of $L$, related to the size of the compact $x^6$
direction. 

We want to know whether there is a stable minimum of the total
potential, and under what conditions this minimum is attained away
from $r=0$. In order to argue for the presence of a stable minimum
at nonzero separation of the brane-antibrane pair, it is sufficient to
show that the potential has an unstable turning point at the
origin. Combined with the attractive behaviour for large
$r$, this suffices to show that the potential develops a stable
minimum somewhere in between.

From Eqns.\energy\ and \smlxpot, we have for $r\ll 1$,
\eqn\smallrv{
V(r) \simeq {1\over g_s (2\pi)^4} {r^2\over L} - Dr^2 }
upto additive constants. Here, $D$ is the positive constant given in
Eqn.\cdvalues. It follows that $V$ has a turning point at the origin
that is unstable (tachyonic) when $L$ is greater than a critical value
$L_c$, namely: 
\eqn\largel{
L > L_c = {1\over g_s (2\pi)^4 D} }
The function $\cF(q)$ defined in Eqn.\fqdef\ tends to the constant
value $-8$ as $t\rightarrow\infty$. Hence an estimate for $D$ can be
made by approximating $\cF$ in the integrand in Eqn.\cdvalues\ by $-16
t^2$ for $0<t<{1\over\sqrt{2}}$ and $-8$ for
${1\over\sqrt{2}}<t<\infty$. With this, we find
\eqn\estimd{
(2\pi)^4 D\sim {16\sqrt{2}\over 2\pi} \sim 3.60 }
so the phase transition takes place at $L_c\sim 0.28\, g_s^{-1}$.

We expect that there will be no loop corrections to the restoring
potential $V^{(1)}$, as this depends only on the D-brane tension which is
unrenormalized. The repulsive potential $V^{(2)}$ will, on the other
hand, receive loop corrections, but they are independent of $L$, and
can be expected to be small for sufficiently small $g_s$. Hence we do
not expect stringy corrections to invalidate the conclusions of this
section.

Some further analysis of this potential can be found in the Appendix. 
\bigskip

\noindent{\bf The T-dual Configuration}

It has been argued that the elliptic configuration involving rotated
NS5-branes is T-dual to the conifold
geometry\refs{\urangaconif,\dmconif}. Above we have studied an
adjacent \ddb4\ pair intersecting with this elliptic
configuration. Hence one may ask what is the precise configuration,
involving a suitable \ddb3\ pair at a conifold, obtained by T-duality
along the compact $x^6$ direction. Such a configuration should
describe a stable non-BPS system exhibiting a phase transition.

As discussed above in the introduction, there is a simpler situation
where the analogous T-duality relation holds: the elliptic model of
two parallel NS5-branes\refs\witfourd, which is T-dual\refs\nsdual\ to
a $Z_2$ ALE geometry. An adjacent \ddb4\ pair in this geometry is dual
to a particular pair of fractional branes at an ALE space\refs\mst. In
this system, the adjacent brane-antibrane pair can separate along the
$(x^4,x^5)$ directions, which lie within the bounding NS5-branes.  In
the T-dual picture the fractional branes live in the 5-plane
transverse to the ALE space, and due to their mutual repulsion, they
separate along the same directions $(x^4,x^5)$, that are transverse
both to their own worldvolume and to the ALE space. In this
discussion, one gets a satisfactory physical picture without having to
take into account the back-reaction of the D-branes on the NS5-branes,
or on the geometry.

For rotated NS5-branes the situation is somewhat different. On the one
hand, the model of a wrapped D4-brane intersecting with rotated
NS5-branes is fairly similar to the one with parallel NS5-branes:
locally there is always a D4-brane ending on a codimension-2 locus
inside an NS5-brane. On the other hand, in the T-dual conifold
geometry we know\refs\kehagias\ that D3-branes completely smoothen out
the conifold singularity: the near-horizon geometry becomes
$AdS_5\times T_{1,1}$. Thus in the latter picture the back reaction of 
the branes on the geometry is qualitatively very important. This can
be traced to the fact that the branes completely fill the space
transverse to the singularity. 

The stable non-BPS configuration discussed in the previous section is
an example of this type. While it can be visualised explicitly in the
brane-construction picture, it is not so easy to describe in terms of
branes at a conifold. For very weak string coupling (and fixed $L$)
the problem is somewhat easier, since in this case the \ddb4\ pair is
aligned. The T-dual configuration will consist of a pair of fractional
branes (more precisely, the first fractional part of a BPS brane,
along with the second fractional part of a BPS antibrane) at the
conifold. As for the case in Ref.\refs\kehagias, here too we expect
the conifold geometry to be smoothed out near the origin. As the
string coupling increases, a phase transition takes place and the
\ddb4\ separates, as discussed above. In this case the T-dual
configuration is harder to visualise. The fractional pair cannot
separate in any direction transverse to the conifold, so it must be
thought of as separating within the conifold
directions. Asymptotically this should look like a
\ddb3\ pair at separated locations away from the conifold
singularity. But close to the origin the behaviour could be more
complicated, with the conifold geometry being replaced by a more
nontrivial one.

Both these situations should be amenable to study as supergravity
solutions. With $N_1$ branes in the first segment and $N_2$ antibranes
in the second, and for sufficiently large $N_1$ and $N_2$, there
should be a trustworthy non-supersymmetric supergravity solution dual
to the RG flow of a non-supersymmetric $SU(N_1)\times SU(N_2)$ gauge
theory. This situation is very similar to the one recently considered
in Ref.\refs{\klebnek,\jpp,\klebtseyt,\ohtatar}, except that these authors
considered supersymmetric configurations with full branes and
fractional branes. If each full brane is replaced by a fractional
brane-antibrane pair (in the sense discussed above) then our desired
configuration is obtained. Since in this process supersymmetry is
completely broken, it remains to be seen whether an explicit solution
can be found. This should be a fascinating direction to explore, as
one would hope to see our phase transition as an instability of the
supergravity solution when some parameter is varied.
\bigskip

\noindent{\bf Summary and Discussion}

We have exhibited a configuration of adjacent D4 and \Dbar4 branes in
type IIA string theory, suspended between relatively rotated
NS5-branes, which corresponds to a stable non-BPS state. A crucial
assumption was that the force between adjacent brane-antibrane pairs
can be estimated using a ``locality'' property, according to which it
originates from the repulsion between the ends of these 4-branes in the
NS5-brane worldvolumes. This repulsion can in turn be computed using
standard orbifold techniques, valid for the model with parallel
NS5-branes which is dual to a $Z_2$ ALE singularity. 

While we do not know at present how to estimate the validity of this
assumption, it is encouraging that it gives a definite and physically
reasonable answer. As we have indicated in the previous section, a
supergravity calculation might be one route to provide an independent 
check of our conclusions.

The $3+1$-dimensional field theory on the common worldvolume in our
brane construction will be a non-supersymmetric, tachyon-free
theory. Because the model is elliptic, it should flow to a CFT. One
can generalise the model to include $N_1$ D4-branes in one segment and 
$N_2$ \Dbar4-branes in the other segment. For $N_1=N_2=N$ this will
again flow to a CFT. Its large-$N$ limit should be interesting.

There are various other generalisations of our model which we have not
discussed here but should be quite straightforward to analyse. This
includes choosing the relative rotation of the NS5-branes to lie
somewhere between 0 and $\pi/2$, introducing more NS5-branes rotated
at various angles\refs\urangaconif, and varying the spacing between
the NS5-branes. One can also study non-elliptic models and incorporate
semi-infinite D4-branes and \Dbar4-branes.

Above the critical radius, our model provides a situation where a
brane and an antibrane are at a finite separation that is calculable
in terms of various parameters including the string coupling and the
radius of a compact direction. Such configurations might perhaps be
useful to construct novel ``brane-world'' type models.
\bigskip

\noindent{\bf Acknowledgements:}

We would like to thank Atish Dabholkar and Sandip Trivedi 
for helpful discussions. We are particularly grateful to Sandip
Trivedi for a careful reading of the manuscript.
\vfill\eject

\noindent{\bf Appendix}

The numerical value of $L_c$ below Eqn.\estimd\ is only approximate,
since we have made a crude estimate for the integral in
Eqn.\cdvalues. An improvement on this estimate can be made by taking
\eqn\betterest{
\eqalign{
\cF(q)~ &\sim~ -16 t^2 + 16 t^4,\qquad t~{\rm small}\cr
\cF(q)~ &\sim~ -8 + 45 q,\qquad t~{\rm large}\cr}}
and then finding an intermediate value of $t$ at which these two
functions match. We find that $t$ is shifted from ${1\over \sqrt{2}}$
to $\sim 0.76$, and the value of $(2\pi)^4 D$ decreases from $3.60$ to
$3.02$.  As a result, $L_c$ moves up to about $0.33 g_s^{-1}$, an
increase of $18\%$. This suggests that at least the order of magnitude
of $L_c$ has been correctly estimated.

One may wonder if the potential has a unique minimum away from 0 for
$L>L_c$ or whether there are several minima, some of them
metastable. For this, it is convenient to make the same approximation
above, but not just for the $r\ll 1$ behaviour. We take the term
$V^{(2)}(r)$ in Eqn.\energy\ and write it as follows:
\eqn\vtwoapprox{
\eqalign{
V^{(2)}(r) &=  - {1 \over 16(2\pi)^4}
\int_{0}^{\infty}{dt \over t^3}
e^{-{8r^{2}t\over \pi}} \cF(q) \cr
& \sim - {1 \over 16(2\pi)^4}
\int_{0}^{1\over\sqrt{2}}{dt \over t^3}
e^{-{8r^{2}t\over \pi}} (-16 t^2)
- {1 \over 16(2\pi)^4}
\int_{1\over\sqrt{2}}^{\infty}{dt \over t^3}
e^{-{8r^{2}t\over \pi}} (-8) \cr}}
The integrals can now be evaluated. It is convenient to rescale the
distance by defining $y={2^{5\over4}\over\sqrt\pi}r$, then we find:
\eqn\vtwoeval{
V^{(2)} \sim {1\over (2\pi)^4} \Bigg( \half (1-y^2) e^{-y^2} +
(1-{y^4\over 2})\Ei(-y^2) - 2\log y - \half - \gamma \Bigg) }
Here, $\Ei$ is the exponential-integral function and $\gamma$ is the
Euler constant. We have dropped an infinite constant associated to the
logarithmic term in the potential, and subtracted a finite constant
$-\half-\gamma$ to make the potential vanish at the origin.

Now we add the first term in Eqn.\energy, which we write:
\eqn\writevone{
V^{(1)} = {1\over g_s(2\pi)^4}
\sqrt{L^{2}+2r^{2}} - L = {\sqrt{\pi}\over 2^{3\over 4}g_s (2\pi)^4}
\Big(\sqrt{\tL^{2}+ y^{2}} - \tL \Big)}
where $\tL= {2^{3\over 4}\over\sqrt\pi}L$, and again a constant has
been subtracted to make the function vanish at the origin. It is now
straightforward to plot $V^{(1)}(y) + V^{(2)}(y)$ for different values
of $\tL$ (Figs.4,5). In these plots we have set $g_s=0.1$. For this
value of $g_s$, the phase transition is at $\tL_c\sim 2.65$. We see
that, at least for the $\tL$ values in the plots, there seems to be
a unique and nonzero minimum when $\tL>\tL_c$.
\bigskip

\centerline{
\epsfbox{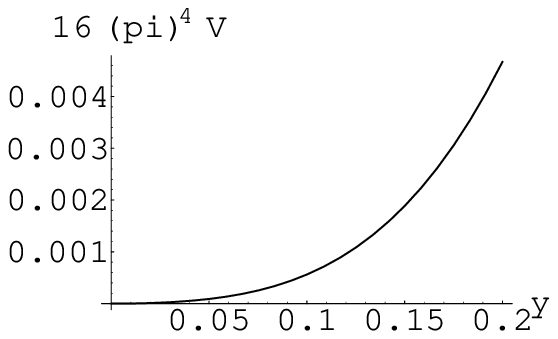}\hfill \epsfbox{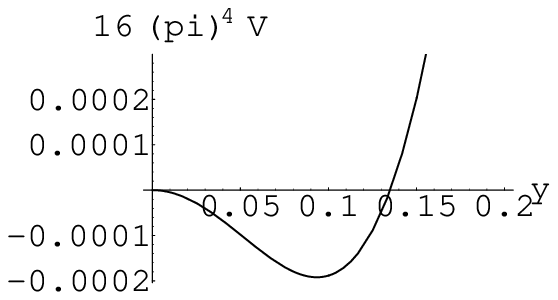}}
\centerline{Fig.4: Potential for $\tL=2.6, 2.7$.}
\bigskip

\centerline{
\epsfbox{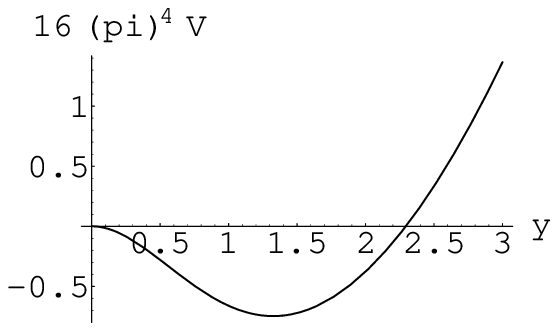}\hfill\epsfbox{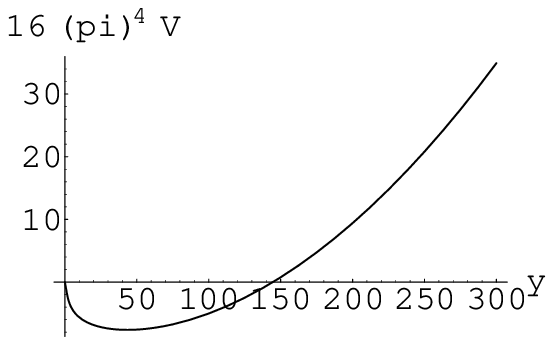}}
\centerline{Fig.5: Potential for $\tL=10,10000$.}
\bigskip

\listrefs

\end